\documentclass[preprint]{revtex4}
\usepackage{graphicx}
\newcommand{\be}{\begin{equation}}
\newcommand{\ee}{\end{equation}}
\newcommand{\bea}{\begin{eqnarray}}
\newcommand{\eea}{\end{eqnarray}}
\newcommand{\lbl}{\label}

\begin{document}

\begin{titlepage}
\title{Scaling behavior of the self diffusion coefficients:\\
dependence on the mass ratio of a binary mixture}

\author{Neeta Bidhoodi, and Shankar P. Das}
\affiliation{School of Physical Sciences,\\
Jawaharlal Nehru University,\\
New Delhi 110067, India.}

\setcounter{equation}{0}

\begin{abstract}
We calculate the self diffusion coefficients $D_S$ for the species
$s=1,2$ of a mixture, and show that a general scaling relation
$D_2{\sim}D_1^a$ with a non universal exponent $a$ holds. The
generalized diffusion coefficients, dependent on the mass ratio
$\kappa=m_2/m_1$ of the constituent particles of the mixture, are
computed using a proper formulation of the self-consistent mode
coupling theory(MCT). The present model for the dense mixture
includes nonlocal and nonlinear effects within the adiabatic
approximation of fast decay of momentum fluctuations compared to
the density fluctuations. The dependence of the slow dynamics near
the characteristic ergodicity-nonergodicity (ENE) transition of
the MCT on $\kappa$ is also studied and a reentrant behavior of of
$D_1$ with respect to $D_2$, in agreement with simulations is
obtained.
\end{abstract}

\pacs{64.70.pm,64.70.qj,61.20.Lc}

\maketitle
\end{titlepage}


The time evolution in the fluctuation of a tagged particle density
$\varrho({\bf x},t)$ in a fluid is a hydrodynamic process
signifying the conservation of the particle. The decay of the time
correlation $\psi(q,t)$ of the local density $\varrho({\bf x},t)$
is set by the corresponding self diffusion coefficient
$D$\cite{boon-yip,mybook}. The latter has been estimated through
extensive experimental and numerical studies \cite{kob}. In a
supercooled liquid a sharp fall in $D$ occurs signifying very slow
single particle dynamics. Binary mixtures are often chosen in
studying glassy dynamics and the behavior of the two
self-diffusion coefficients $D_s$ corresponding to the respective
species $s=1,2$ is an issue of much interest. Both $D_1$ and $D_2$
characterize the mixing process in the fluid  and computer
simulations\cite{hl-prl} show that their behavior are strongly
dependent of the mass ratio of the constituent particles of the
mixture. Understanding this dependence of the slow dynamics
involving $D_1$ and $D_2$ on the characteristic properties of the
mixture, like size or mass asymmetry of its particles is the
motivation of the present work.

A microscopic model for the correlated dynamics of the particles
of a dense binary mixture has been formulated\cite{uh-pre,bin1}
using the self-consistent mode coupling approach\cite{rmp}. The
mode coupling theory (MCT) in its simplest
form\cite{gtze,balatz,bosse} predicts a ENE transition in the
mixture beyond a critical packing and has been used extensively in
studying slow dynamics \cite{rabani}. The model which takes in to
account the proper conservation laws for the two component system
shows that the transition point is dependent on the mass ratio
\cite{uh-jsp} of the constituent particles. This dependence of the
dynamics on the mass ratio has been seen in simulations
\cite{hl-prl}. By applying the so called adiabatic approximation
in the MCT model, we presented \cite{bin2} the mechanism for
producing very slow dynamics of a tagged particle in an one
component liquid. In the present work we show how the asymmetric
mass ratio influences the self-diffusion coefficients $D_1$ and
$D_2$ of the two species of the mixture. The findings of our model
are in agreement with simulation results.


The MCT is formulated primarily in terms of correlation of a small
set of collective densities. Let us first consider the mechanism
for producing very long relaxation time for the tagged particle
correlation $\psi(q,t)$ in a one component fluid due to mode
coupling effects. The MCT is formulated in this case in terms of
the Laplace transform of the tagged particle correlation
$\psi(q,t)$,\be \label{ocp-difu}
\psi(q,z)={[z+iq^2D(q,z)]}^{-1}~,~~\ee where the generalized
memory function $D(q,z)$ is written in the inverse form
\cite{kawa_1995,bin2} as $D^{-1}(q,z)=D_0^{-1}+\Delta_{mc}(q,z)$.
Here $D_0$ is the bare self-diffusion coefficient and
$\Delta_{mc}(q,z)$ is the mode coupling contribution. At the one
loop order, the mode coupling part is obtained using the adiabatic
approximation as
\be \label{mem-one} \Delta_{mc}(q,z) =n_0\int\frac{d{\bf
k}}{(2\pi)^3}\frac{u^2}{k^2}V^2(q,k) {\cal C}(k) \psi(q-k,t)~~,
\ee
where $u$ is the cosine of the angle between ${\bf q}$, $\bf k$.
The quantity ${\cal C}(k,t)$ is related to the collective density
correlation function $\phi(q,t)$ as ${\cal
C}(k,t)=\{{c}(k)\phi(k,t){c}(k)\}$, where the weight factor $c(k)$
is Ornstein-Zernike two point direct correlation function
normalized with a factor of average number density $n_0$. The
vertex function $V(q,k)$ in the mode coupling integral of Eqn.
(\ref{mem-one}) is obtained as, $V(q,k)={k_BT}/ (D_0(q)L_0(k))$,
where $L_0$ and $D_0$ respectively denote the corresponding bare
viscosity and the bare self diffusion coefficient. Taking the
transport coefficients for their respective hydrodynamic values,
the vertex factor is like a characteristic length scale similar to
that defined with respect to Stokes-Einstein relation
\cite{se-reln}. By transforming Eqns. (\ref{ocp-difu}) and
(\ref{mem-one}) to the time domain we obtain a coupled  set of
nonlinear integro-differential equations for $\psi(q,t)$ for
different values of wave vector $q$. These equations contain time
convolutions with the memory function ${\Delta_\mathrm{mc}}$. The
latter is expressed in the mode coupling formulation in terms
${\cal C}$ which involves the correlation function $\phi(k,t)$ for
the collective density. Thus to evaluate tagged particle
correlation $\psi(q,z)$ and hence the self diffusion coefficient
$D(q,z)$, we need to obtain independently the total correlation
$\phi(q,z)$.

For the two component system, the relevant densities are the
partial (mass) densities $\{\rho_1,\rho_2\}$ of the two species
and the total momentum density ${\bf g}$. They respectively
correspond to the conservation of mass of individual species and
total momentum in the mixture. Linear combinations of $\rho_1$ and
$\rho_2$ constitute an equivalent set of collective modes
$\{\rho,c,{\bf g}\}$, where $\rho=\rho_1+\rho_2$ and
$c=x_1\rho_2-x_1\rho_1$. $x_s$ is the fraction particles of
species $s$. The theoretical analysis presented below, broadly
consists of two parts. In the first part, we present the equations
for the time evolution of the tagged particle correlations
$\psi_s(q,t)$ for the mixture in terms of the corresponding memory
functions $\Delta_\mathrm{mc}^s$ respectively for $s=1,2$. We
approximate the respective memory functions in a self-consistent
form in terms of convolutions of $\psi_s$ and the correlations of
collective correlation functions. The latter are denoted as
$\{\phi_{\alpha\sigma}\}$, with $\alpha,\sigma\in\{\rho,c\}$. In
constructing these memory functions $\Delta_\mathrm{mc}^s$, we
follow the corresponding results of MCT for a one component system
obtained using the adiabatic approximation. In the second part of
our analysis, we focus on the calculation of the collective
correlation functions$\{\phi_{\alpha\sigma}\}$, with
$\alpha,\sigma\in\{\rho,c\}$ needed for evaluating the self
consistent expressions for the memory functions obtained in the
first part. The self diffusion coefficients $D_s$, its dependence
on the characteristic properties like mass ratio, size ratio, etc.
are then computed.

Generalizing the results for the one component fluid, the
corresponding equation of motion for the tagged particle
auto-correlations $\psi_s(q,t)$ of species $s$ (for $s=1,2$) is
obtained.
\be \label{self:binary} \dot{\psi}_s (q,t)+D_0^s q^2 \psi_s(q,t) +
D_0^s \int_0^t ds~ \Delta_{mc}^s(q,t-s)\dot{\psi}_s(q,s)=0~~, \ee
where $D^s_0$ denotes the bare diffusion coefficient for species
$s$ \cite{marchetti}. The memory function $\Delta_{mc}^s(q,t)$
corresponding to $s$-th species is written in the form:
\be \label{self:mct} \Delta_{mc}^s(q,t) = n_0\int \frac{d{\bf
k}}{(2\pi)^3}\frac{u^2}{k^2}V^2_s(q,k)~ {\cal
C}_s(q,t)\psi_{s}(q-k,t)~~~~. ~\ee
The vertex function $V_s(q,k)$ in the above equation is a
generalization of the corresponding quantity $V(q,k)$ defined in
Eqn. (\ref{mem-one}), with the self-diffusion coefficient $D_0$
being replaced by the corresponding $D^s_0$. The collective part
of the correlation, ${\cal C}_s(k,t)$  in Eqn. (\ref{self:mct}),
is obtained by generalizing the ${\cal C}(k,t)$ in the right hand
side of Eq. (\ref{mem-one}) for the one component case. The
$ss$-th element of the corresponding $2\times{2}$ matrix is
written as
\be \label{col-s} {\cal C}_s(q,t) = \Big[{\bf c}^T(k)\star
\Phi(q,t) \star {\bf c} (k) \Big ]_{ss}~~. \ee
$\Phi$ and ${\bf c}$ are the respective generalizations for the
two component case \cite{KB} of the normalized density correlation
$\phi(k,t)$ and Ornstein-Zernike direct correlation functions
$c(k)$ for the one component fluid. The superscript $T$ in ${\bf
c}^T$ stands for transpose of the $2\times{2}$ matrix ${\bf c}$ in
in terms of species index $s,s'$ etc. Similarly the correlation
function matrix $\Phi$ is defined as $\Phi_{ss'}(k,t) =
\{\sqrt{S_{ss}(k)S_{{s'} {s'}}(k)}~ \phi_{ss'}(k,t)\},~$ where,
$\phi_{ss'}$ is the equilibrium correlation of fluctuations of the
partial densities $\rho_s(k,t)$ and $\rho_{s'}(-k,t)$:
$\phi_{ss'}(k,t)=\langle{\delta\rho_s}(k,t)\delta\rho_{s'}(-k,0)\rangle
/{\sqrt{N_sN_{s'}}}~~$. The matrix $S_{ss}(k)$ is the equal time
correlation function $\phi_{ss'}(k,t=0)$ ($s=1,2$).
${\cal C}^s(q,t)$ is the key quantity through which the collective
modes of the mixture couple to the single particle modes in the
respective memory function $\Delta_{mc}^s$ (for $s=1,2$). In
writing Eqn. (\ref{self:binary}) we make the plausible assumption
that the dynamics of the {\em single} particle modes \{$\psi_s$\}
of the respective species $s=1,2$ do not influence each other, and
hence the time evolution of $\psi_1$ and $\psi_2$ are not linked.
Each of the single particle correlations couple only to the
collective modes in the mixture.


The coupled set of Eqns. (\ref{self:binary})-(\ref{self:mct}) are
solved self-consistently  on a wave vector grid to obtain the self
correlation functions $\psi_s(q,t)$ (for $s=1,2$). The memory
functions $\Delta^s_{mc}$ ($s=1,2$) in the respective equations
for $\psi_s$ ($s=1,2$) depend on ${\cal C}_s$ and hence on
collective correlation functions $\{\phi_{\alpha\sigma}\}$. The
latter are obtained from MCT for the binary mixture
\cite{bin1,bin2} which follows from the corresponding equations of
fluctuating nonlinear hydrodynamics. The renormalized theory which
takes in to account the effect of the nonlinearities is
constructed with the standard Martin-Siggia-Rose field theoretic
approach. Using the available fluctuation dissipation relations
between the correlation and response functions, we obtain the
equations for the time evolutions of $\phi_{\alpha\sigma}$ with
$\alpha,\sigma\in\{\rho,c\}$. The dynamics is considered here in
the so called adiabatic approximation in which the momentum
fluctuations decays out much faster than the density fluctuations.
These details are provided in the Appendix \label{app}.

The single particle dynamics becomes slaved to collective
correlation in the adiabatic approximation as is evident from the
coupled set of Eqns.(\ref{self:binary}) and (\ref{self:mct}). The
renormalized self-diffusion coefficient for species $s$ ($s=1,2$)
is obtained as $D_s=D^s_0/(1+\tilde{\Delta}_s)$ where
$\tilde{\Delta}_s$ is the static limit of the corresponding memory
function $\Delta^s_{mc}$, {\em i.e.}, we write
$\tilde{\Delta}_s=D^s_0\int_0^\infty\Delta^s_{mc}(0,t)dt$. The
memory function $\Delta^s_{mc}(0,t)$ is expressed in terms of
correlation functions of collective modes $\{\rho,c\}$ when the
adiabatic approximation \cite{bin2} is applied in the overdamped
limit. Hence beyond the ENE transition $\Delta^s_{mc}(0,t)$ is
constant over long times and we have
$\tilde{\Delta}_s\rightarrow\infty$. Hence $D_s\rightarrow{0}$.
Single particle diffusion constants $D_1$ and $D_2$ vanish only
with application of the adiabatic approximation. The location of
the characteristic ENE transition is obtained from the long time
behavior of the correlation of collective variables in the fluid.
Beyond the ENE transition point
$\phi_{\alpha\sigma}(q,t\rightarrow\infty)=f_{\alpha\sigma}(q)$
are non zero and this causes $\psi_s(q,t\rightarrow\infty)=f_s(q)$
(for $s=1,2$) being nonzero. $\{f_{\alpha\sigma}(q)\}$ and $f_s$
termed as nonergodicity parameters.

\begin{figure}[htbp!]
\centering
\includegraphics[width=0.6\textwidth]{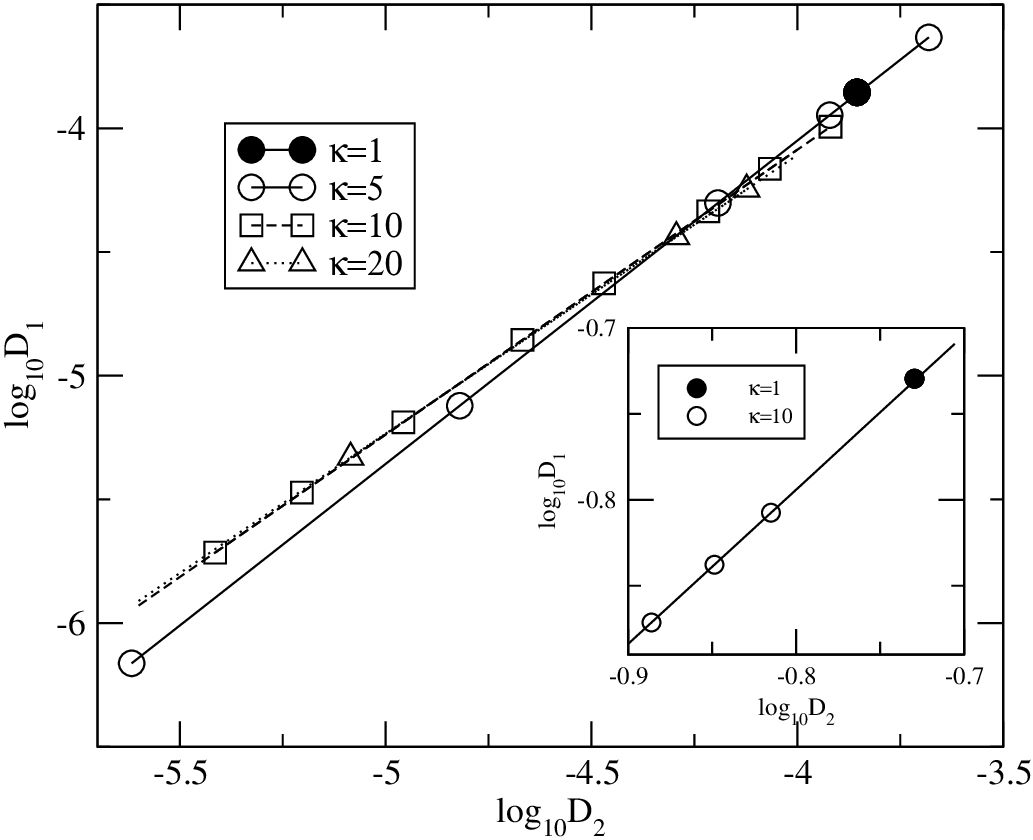}
\caption{Self-diffusion coefficients (expressed in units of
$\sigma_2/\sqrt{\beta m_1}$) obtained from MCT for a hard sphere
mixture with size ratio $\alpha=1$ and respective mass ratios
$\kappa=5$(circle),$10$(square),$20$(triangle). The plots follow
the scaling relation $D_2={\vartheta}D_1^a$ with $a=1.30$, $1.15$,
and $1.12$ respectively. Inset shows simulation results of a
truncated and shifted Lennard-Jones mixture for $\alpha=1$ and
$\kappa=10$ (reproduced from Ref. \cite{sim-pre}). The straight
lines for different $\kappa$ values are shifted along the
horizontal axis so as to pass through the common point of the
filled circle representing the one component system. The shifting
is accounted for by the respective amplitude factors
$\vartheta(\kappa)$ in the scaling relation.} \label{fig1}
\end{figure}

\begin{figure}[htbp!]
\centering
\includegraphics[width=0.6\textwidth]{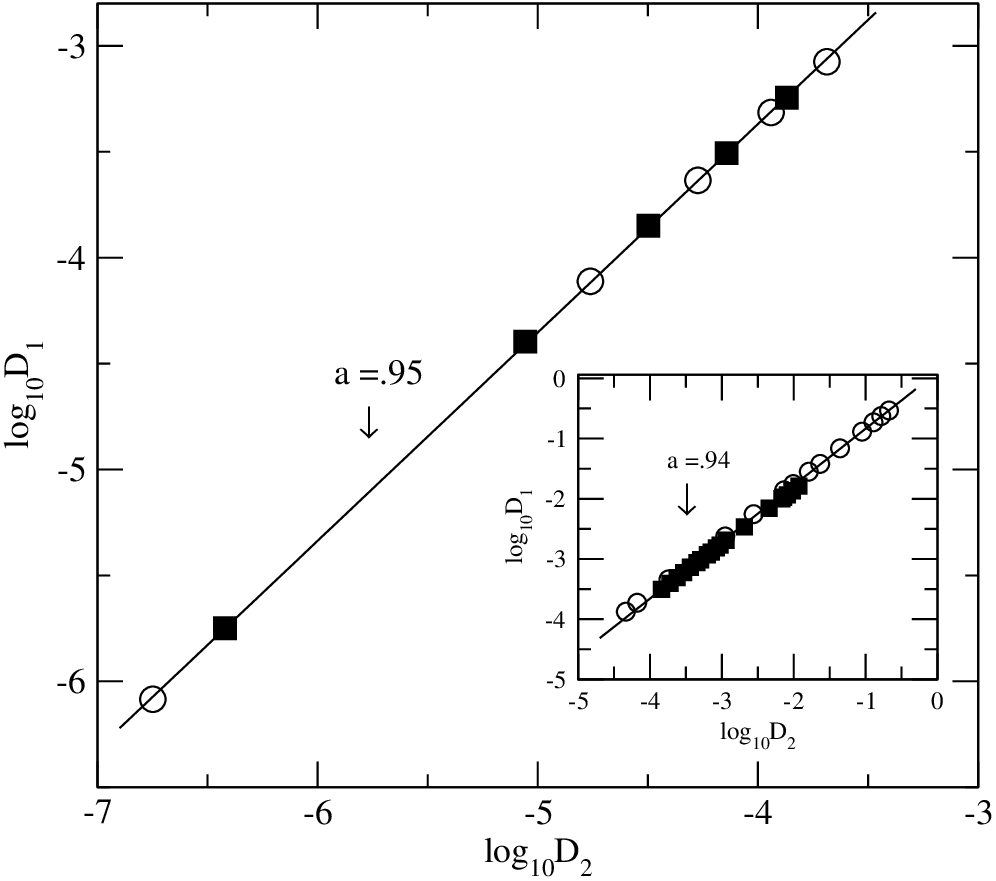}
\caption{Self-diffusion coefficients (expressed in units of
$\sigma_2/\sqrt{\beta m_1}$) obtained from MCT for a binary
mixture of hard spheres of size ratio $\alpha=1.2$, mass ratio
$\kappa=1$ and relative abundance $x=.8$(circle),$.5$(square). The
data corresponds to the total packing fraction $\eta$ range
$\{.500$-$.520\}$. Inset shows simulation results (reproduced from
Ref. \cite{sastry}) with circle(square) for a Kob-Anderson(Square
Well) mixture with size ratio $\alpha=1.25(1.20)$, and mass ratio
$\kappa=1(1)$. The range of temperatures, scaled with respect to
the corresponding potential depth $\epsilon$, are $\{0.46,
6.0\}$($\{0.31-10\}$). The scaling exponents are $a=.95$(main) and
$.94$ (inset).}\label{fig2}
\end{figure}

In the following, we present results for the dynamics of the hard
sphere mixture from two broad aspects. First, we study the
self-consistent MCT predictions for single particle correlations
$\psi_s(q,t)$, and hence the self diffusion coefficients $D_s$ for
$s=1,2$, when the mixture is in the ergodic phase. We study the
dependence of MCT predictions for $D_1$ and $D_2$ on the mass
ratio $\kappa=m_2/m_1$, size ratio $\alpha=\sigma_2/\sigma_1$ of
the particles of the two species of the mixture, and the relative
abundance $x=N_2/N$ of species $2$. Second, we consider the
characteristic ENE transition point of MCT and study how the
critical value of total packing fraction $\eta_c$ changes with
$\kappa$, $\alpha$ and $x$.
In Fig. \ref{fig1} we display $D_1$ and $D_2$ on a log-log plot
for binary mixture with size ratio $\alpha=1.2$ three different
values of the mass ratio. The straight line fits indicate the two
self diffusion coefficients $D_1$ and $D_2$ are simply related as
$D_1={\vartheta}D_2^a$ with the $a$ and $\vartheta$ being
non-universal. As $\kappa$ increases the slope $a$ converges to a
value close to $1.12$. The study is done for a fixed value of the
total packing fraction $\eta=.500$ of the mixture. Thus the
variation of the diffusion constants along each curve, corresponds
to a range of $x$. For lines shown in Fig. \ref{fig1}
corresponding to $\kappa=5,10$, and $20$ the respective range of
$x$ values are $[.75-.85]$, $[.85-.90]$. In the inset of Fig.
\ref{fig1} we show diffusion data from simulation of mixture of
soft spheres \cite{sim-pre} having size ratio $\alpha=1.0$, and
mass ratio $\kappa=10$. The diffusion data shown are at fixed
packing $\eta=.313$ and temperature $T^*=1.05$, with the relative
abundance $x$ varying over the range $[.05-.2]$. The straight
lines for different $\kappa$ values in Fig. \ref{fig1} are shifted
along the horizontal axis so as to make each pass through the
common point of the filled circle representing the one component
system ($\alpha=1$,$\kappa=1$). This shifting accounts for the
amplitude factor $\vartheta(\kappa)$ in the scaling relation. For
the three curves corresponding to $\kappa=5,10,$ and $20$, shown
in Fig. \ref{fig1} we obtain $\vartheta=15.0, 3.5,$ and $2.8$. For
the simulation data in the inset of Fig. \ref{fig1} the same
scaling holds with $\vartheta=.87$.
In Fig. \ref{fig2} we show the results of the MCT model for the
diffusion constants for different packing fractions $\eta$ at a
fixed value of relative abundance $x$. Mass ratio  and size ratio
characterizing the mixture are kept same for each set of data. The
corresponding range of $\eta$ is $\{.500-.520\}$. The diffusion
coefficients follow the behavior $D_2{\sim}D_1^a$ with $a=.96$. In
the inset of Fig. \ref{fig2} we show similar behavior seen in
simulations of two different kinds of interaction potentials,
namely, binary Lennard-Jones and Square well interaction
\cite{sastry}.

\begin{figure}[htbp!]
\centering
\includegraphics[width=0.6\textwidth]{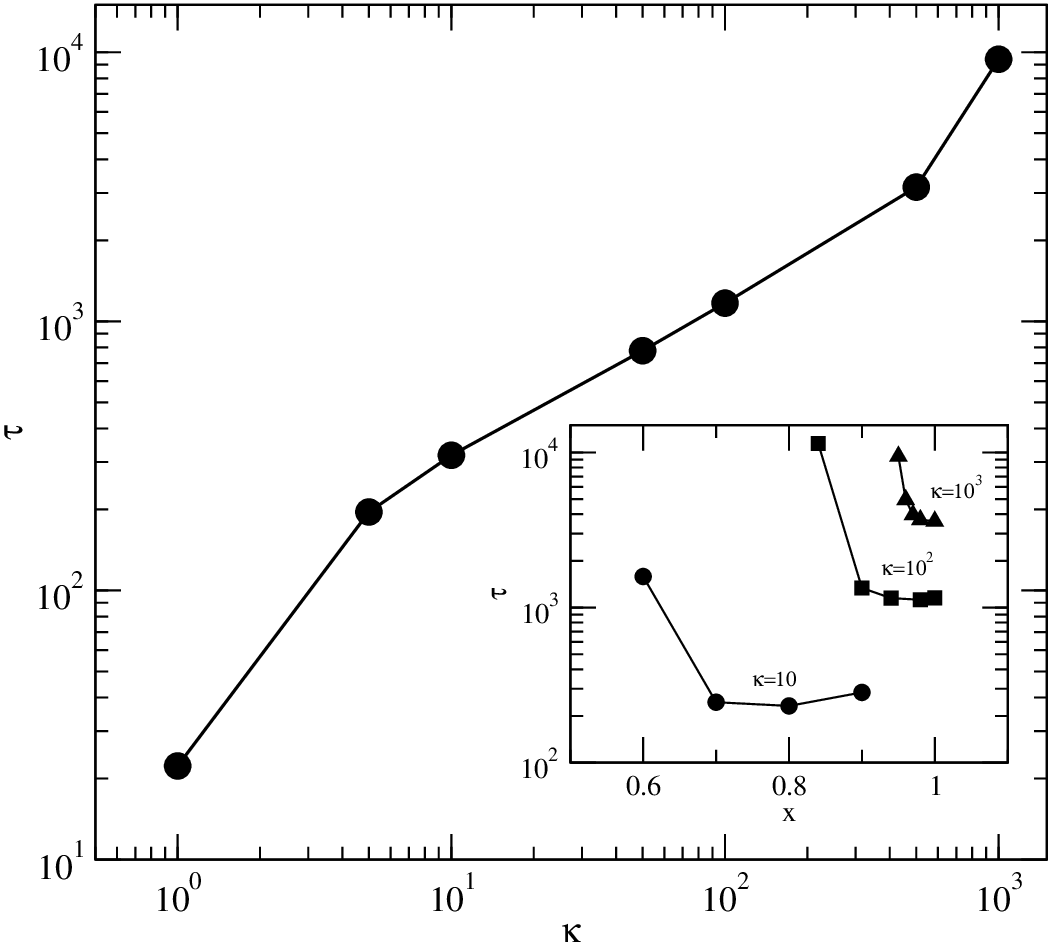}
\caption{Relaxation time $\tau$ for the total density correlation
$\psi_{\rho\rho}$ near the structure factor peak vs. mass ratio
$\kappa$ for a mixture of hard spheres, with size ratio
$\alpha=10$, relative abundance $x=.95$ and total packing fraction
$.5$. Inset: $\tau$ vs. $x$ for size ratio $\alpha=10$. The curves
are labeld with corresponding mass ratio $\kappa$}\label{fig3}
\end{figure}

Binary mixtures are very useful systems for studying glassy
dynamics in computers, since they do not readily crystallize. An
important issue in this regard is ease of equilibration for the
mixture and this is controlled by its relaxation time. The
dependence of the relaxation time $\tau$ on the mass ratio
$\kappa$ of the mixture is shown in Fig. \ref{fig3}. The
relaxation time $\tau$ is estimated from the time dependence of
the decay of total density correlation $\psi_{\rho\rho}(q,t)$ at
the peak of the corresponding structure factor $q_m\sigma_2=7.0$.
The parameters of the mixture are kept fixed at $x=.95$
$\alpha=10$ and total packing fraction $\eta=.500$. As in Fig.
\ref{fig1} and Fig. \ref{fig2}, we study behavior of $\tau$ with
change of either one of $x$ or $\kappa$, while the other is kept
fixed. Generally as the bigger sized particles get more massive,
relaxation slows down. On the other hand for the mixture of same
species, {\em i.e.}, constant $\kappa$ and $\alpha$, relaxation
slows down as the mixture becomes more asymmetric in the relative
abundance of the two species.

\begin{figure}[htbp!]
\centering
\includegraphics[width=0.6\textwidth]{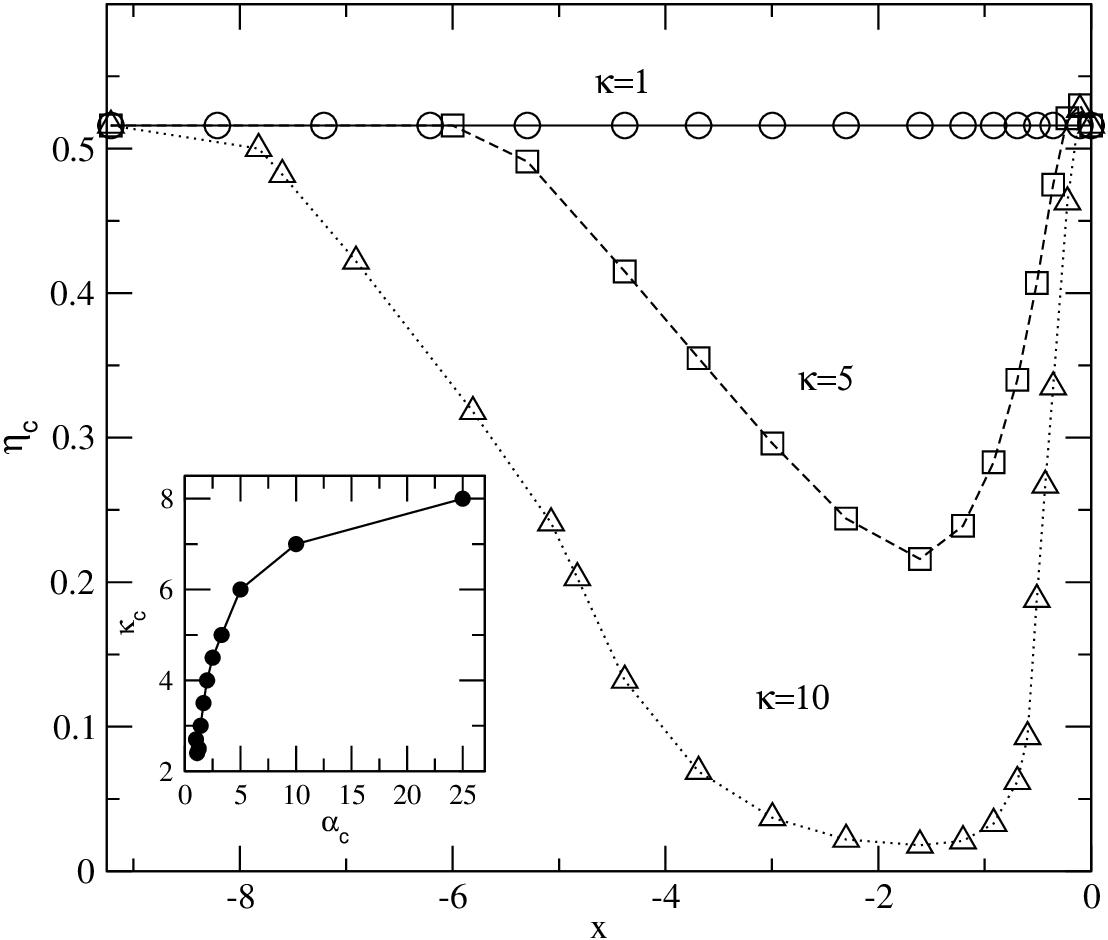}
\caption{The critical packing fraction $\eta_c$ corresponding to
the ENE transition of MCT vs. the relative abundance $x$ of a hard
sphere mixture having size ratio $\alpha=1$, and mass ratio
$\kappa=$1(circle), 5(squares), and 10(traingles). Inset shows for
a mixture with packing $\eta=.5$ and $x=.5$, the pairs of values
for mass ratio $\kappa_c$ and corresponding size ratio $\alpha_c$
at the ENE transition of MCT.}\label{fig4}
\end{figure}

\begin{figure}[htbp!]
\centering
\includegraphics[width=0.6\textwidth]{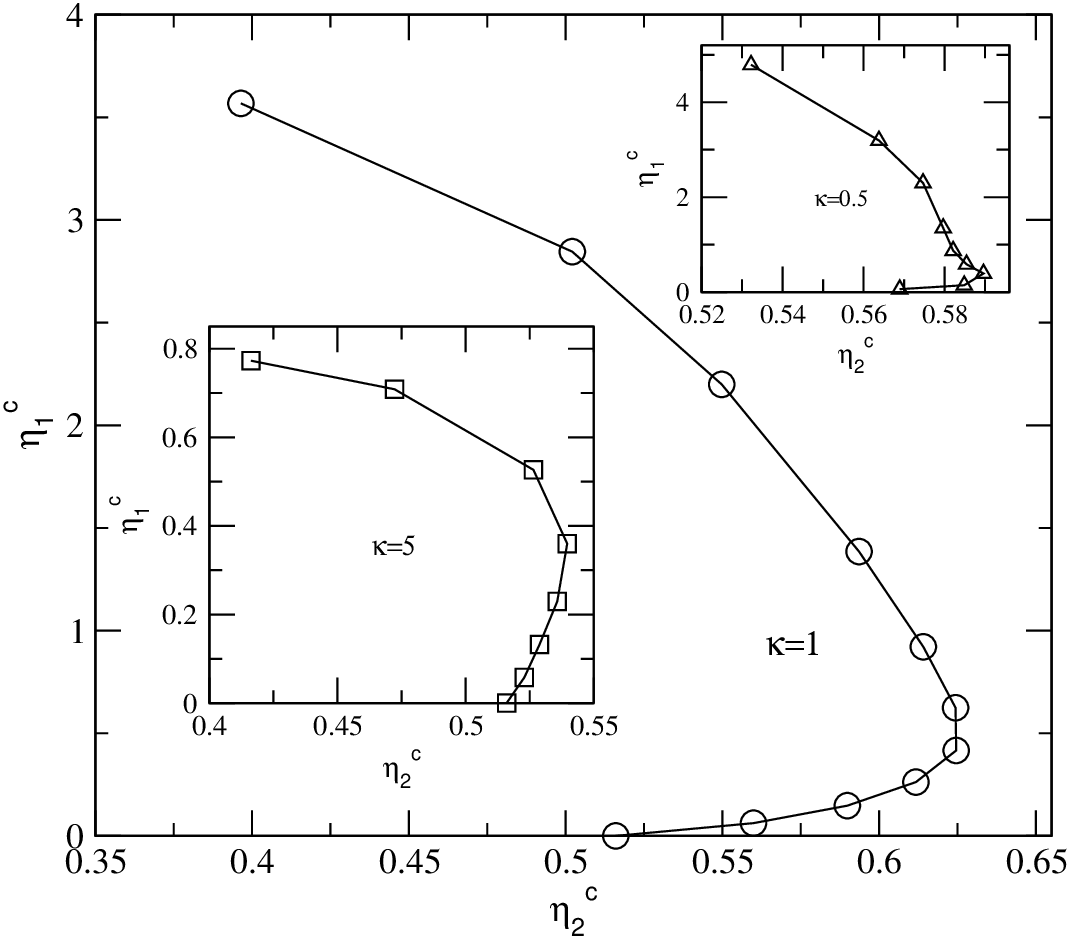}
\caption{The respective packing fractions of the two species at
the ENE transition points of MCT, $\{\eta_1^c,\eta_2^c\}$ for a
hard sphere mixture of size ratio $\alpha=10$ and mass ratio
$\kappa=1$(main), $.5$ (upper inset), and $5$ (lower inset).
Reentry of the critical behavior with respect to $\eta_2^c$ is
seen here since at a fixed $\eta_2^c$, two possible values of
$\eta_1^c$ occur.}\label{fig5}
\end{figure}

Finally we study the role of the mass ratio $\kappa$ on the
possible ENE transition in the mixture. The location of the
transition point, in terms of the critical value of the total
packing fraction $\eta=\eta_c$, vs. relative abundance $x_2$ is
shown for three different $\kappa$ in Fig. \ref{fig4}. The size
ratio for all the curves are fixed at $\alpha=1$. For the
$\kappa=1$ we obtain for all $x$, the one component result showing
that $\eta_c=.516$ independent of $x$. For $x=1$ or $x=0$, we have
an one component system and hence the $\eta_c$ for the hard sphere
system is fixed at $.516$. With increasing $\kappa$ however the
ENE transition can occur at much lower packing at intermediate
values of $x$. This is shown Fig. \ref{fig4}. In Ref.
\cite{hl-prl} the reentrant nature of the ENE transition was
observed from simulations of a two component mixture. The system
studied in Ref. \cite{hl-prl} is not a hard sphere system. However
we also observe similar behavior for a hard core system from our
model equations. In Fig. \ref{fig5} we plot the pairs of partial
packing fraction values $\{\eta_1^c,\eta_2^c\}$ corresponding to
the ENE transition of the MCT. Mixtures with three different
$\kappa$ values, keeping the size ratio fixed at $\alpha=1$, are
considered. In each case as the relative abundance changes the ENE
transition occurs for different $\{\eta_1^c,\eta_2^c\}$ pairs. The
nature of the criticality curves show that for certain $\eta_2$
values ( depending on $\kappa$ ) reentry in to the liquid phase
occurs as $\eta_1$ is increased.

The mass ratio dependence of the dynamics which was absent in
earlier MCT models, is seeded in the basic formulation of the
present model. In earlier MCT formulations, the mass densities
\{$\rho_1$,$\rho_2$\}, and the momentum densities \{${\bf
g}_1$,${\bf g}_2$\} of each component of the mixture  were treated
as slow variables. With the four ``slow" variables $\{\rho_s,{\bf
g}_s\}$, for $s=1,2$ the key nonlinearities in the corresponding
equations of FNH (shown in the Appendix \label{app}) are
$\rho_s{\nabla_i}(\delta{F_U}/\delta{\rho_s})$. The mass
dependence of the drops out in the corresponding formulation of
the MCT. In subsequent works \cite{uh-jsp,hl-prl} instead of $\{
{\bf g}_1,{\bf g}_2\}$, only the conserved {\em total} momentum
density ${\bf g}={\bf g}_1+{\bf g}_2$ was taken as a slow mode.
The corresponding set of Langevin equation for the mixture which
gives rise to the mode coupling terms has the relevant non
linearity in the equation for ${\bf g}$ of the form
$\rho{\nabla_i}(\delta{F_U}/\delta{\rho})$,
$c{\nabla_i}(\delta{F_U}/\delta{c})$. This term is {\em dependent}
on the mass ratio of the two species and hence the latter affects
the dynamics. NB acknowledges CSIR, India, and SPD acknowledges
the JC Bose fellowship of DST, for financial support.


\setcounter{equation}{0}

\appendix{
\section*{Appendix}
\label{app}

\noindent The coupled dynamics of the time correlation functions
$\phi_{\alpha\sigma}(q,t)$ of the {\em collective} modes
$\alpha,\sigma\in\{\rho,c\}$ is controlled by the equations of
fluctuating nonlinear hydrodynamics (FNH) for a mixture. These
equations signify the basic conservation laws for the system and
involve nonlinear couplings of the collective modes. We explain
here the steps of calculating the correlation functions of the
collective densities for a binary mixture of $N_s$ identical
particles of species $s$ having mass $m_s$ for $s=1,2$
respectively. The concentration of the species $s$ is $x_s=N_s/N$
and $N=N_{1}+N_{2}$ is the total number of particles. The
individual mass densities $\rho_s$ and the momentum densities
${\bf g}_s$ for the species $s$ are respectively defined in terms
of microscopic phase space variables as follows,
\bea \label{rhos-def}\rho_{s}({\bf x},t) &=&
m_{s}\sum_{i=1}^{N_{s}}\delta({\bf x}-
{\bf R}_{s}^{i}(t)), \\
\label{gs-def} {\bf g}_{s}({\bf x},t) &=& \sum_{i=1}^{N_{s}}{\bf
p}_{s}^{i}\delta({\bf x} -{\bf R}_{s}^{i}(t))~~. \eea
The phase space coordinates of position and  momentum of the
$i$-th particle of the species $s$ are denoted as
$\{R_s^i(t),P_s^i(t)\}$. The individual mass densities,
respectively denoted as $\rho_s({\bf x},t)$ are microscopically
conserved. The individual momentum densities ${\bf g}_1$ and ${\bf
g_2}$ are not conserved but total momentum density defined as,
$ {\bf g}({\bf x},t)={\bf g}_{1}({\bf x},t)+{\bf g}_{2}({\bf
x},t)$
is conserved. In the following we consider the dynamics in terms
of a different set of conserved variables: the total mass density
$\rho({\bf x},t)$, total momentum density ${\rm g}({\bf x},t)$,
and the concentration variable $c({\bf x},t)$ \cite{cohen}. The
mass and concentration densities are defined as follows:
\bea \lbl{totden-def}
\rho({\bf x},t) &=& \rho_{1}({\bf x},t)+\rho_{2}({\bf x},t), \\
\lbl{conc-def}
 c({\bf x},t)&=& x_{2}\rho_{1}({\bf x},t)-x_{1}\rho_{2}({\bf
 x},t).
\eea
We define the fluctuations of $\rho$ and $c$ respectively as
$\delta\rho=\rho-\rho_0$ and $\delta{c}=c$, since the average of
$c$ is zero when we consider the mass ratio of the constituent
particles to be unity.

The generalized Langevin equations for the respective coarse
grained densities $\{\rho({\bf x},t),{\bf g}({\bf x},t),c({\bf
x},t)\}$ for a binary mixture are obtained following standard
procedures \cite{mybook}.
\bea && \label{cont-eqn}
\frac{\partial\rho}{\partial t}+\nabla . {\bf g}=0, \\
&& \label{momt-eqn} \frac{\partial {\rm g}_i}{\partial t}+ {\bf
\nabla}_j \left [ \frac{{\rm g}_i{\rm g}_j}{\rho} \right
]+\rho\nabla_{i} \frac{\delta
F_{U}}{\delta\rho}+c\nabla_{i}\frac{\delta F_{U}}{\delta c} +
L^0_{ij}\frac{{\rm g}_j}{\rho}
=\theta_{i},\\
&& \label{conc-eqn} \frac{\partial c}{\partial t}+{\bf
\nabla}\cdot \left [ c\frac{\bf g}{\rho} \right ]
+\gamma_{cc}\nabla^2 \frac{\delta F_U}{\delta c}=f_c. \eea
$F_U$ is the so called potential  part of the free energy
functional $F$ which is expressed as \be F[\rho,{\bf
g},c]=F_K[\rho,{\bf g}]+F_U[\rho,c], \ee
where the kinetic part $F_K$ (dependent on the current density
${\bf g}$) is computed from the microscopic Hamiltonian for the
$N$ particle mixture \cite{uh-pre}.
$F_U[\rho,c]$ is taken here as a quadratic functional of the
fields $\rho$ and $c$, and is related to the structure of the
liquid. This is expressed in terms of the corresponding direct
correlation functions $\{c_{\rho\rho},c_{\rho{c}},c_{cc}\}$
defined in the Ornstein-Zernike relations. $L_{ij}^0$ represents
the matrix of bare or short time viscosities while $\gamma_{cc}$
links to the bare inter-diffusion coefficient for the mixture.
These two dissipative coefficients are related to the correlation
of the Gaussian noises respectively in Eqs. (\ref{momt-eqn}) and
(\ref{conc-eqn}) through standard fluctuation-dissipation
relations \cite{bin1}.

Effects of these nonlinearities on the correlation functions are
obtained using a Martin-Siggia-Rose (MSR) field theory. In this
formulation the renormalized transport coefficients are obtained
in a self consistent form\cite{bin1,bin2}. Applying the available
fluctuation dissipation relations \cite{ABL} in the model and
introducing the the intermediate kernel function $\Xi(q,t)$, the
equations of motion for correlation functions
$\phi_{\alpha\sigma}$ are obtained as follows:
\bea \label{tde:rr}&& \dot{\phi}_{\rho\rho}(q,t)
+\Omega_q^2\int_{0}^{t}ds~\Xi(q,t-s)\tilde{\phi}_{\rho\rho}(q,s)=0~,\\
\label{tde:rc}&& \dot{\phi}_{\rho c}(q,t)+ \Omega_q^2
\int_{0}^{t}ds~\Xi(q,t-s)\tilde{\phi}_{\rho c}(q,s)=0~,\\
\label{tde:m}&&\dot{\Xi}(q,t)+ \Omega_q^2\left\{L_0(q) \Xi(q,t)+
\int^{t}_{0}ds~ L(q,t-s)\Xi(q,s)\right\}=0~,\\
\label{tde:cc}& & \dot{\phi}_{cc}(q,t)+\nu_0q^2\tilde{\phi}_{c
c}(q,z)+ \int_{0}^{t}ds~M(q,t-s)\dot{\phi}_{cc}(q,s)=0~.\eea
The quantity $\Omega_q=qc_0$ represents a characteristic frequency
scale for the mixture in terms of the sound speed $c_0$ and we
have defined the dressed correlations:
$\tilde{\phi}_{\rho\alpha}=\phi_{\rho\alpha}-\chi\phi_{c\alpha}$
and $\tilde{\phi}_{cc}=\phi_{cc}- \chi\phi_{\rho c}$, with $\chi
=\chi_{\rho c}/\sqrt{\chi_{\rho\rho}\chi_{cc}}$,
$\alpha{\in}\{\rho,c\}$.
$L_0(q)$ is the bare longitudinal viscosity to which a mode
coupling contribution $L(q,t)$ is added. Similarly the inverse of
the inter-diffusion coefficient
of the mixture includes a mode coupling contribution $M(q,t)$. In
writing Eqns. (\ref{tde:rr})-(\ref{tde:cc}), the presence of
$1/\rho$ nonlinearities \cite{sdd_com,DM09} in the FNH equations
for a mixture have been ignored. The memory functions $M(q,t)$ and
$L(q,t)$ - {\bf mode-coupling expressions} in terms of
correlations involving $\{\rho,c\}$. In the adiabatic
approximation relevant for the glassy dynamics, the simplest mode
coupling expressions for $L(q,t)$ and $M(q,t)$ are obtained as
bilinear products involving correlation functions
$\phi_{\alpha\sigma}$ of the fields $\{\rho,c\}$. These coupled
set of integro-differential equations are solved with the initial
conditions $\psi_{\rho\rho}(q,0)=1$, $\psi_{\rho c}(q,0)=\chi$ and
$\psi_{cc}(q,0)=1$. The coupled dynamics of the correlations of
the $\{\rho,c\}$ fields is strongly dependent on the mass ratio
$\kappa$ of the mixture since the effective nonlinearities and
hence the memory functions in Eqns. (\ref{tde:rr})-(\ref{tde:m})
involves $\kappa$ in the present model. Beyond a ENE transition
point
$\phi_{\alpha\sigma}(q,t\rightarrow\infty)=f_{\alpha\sigma}(q)$
are non zero. $\{f_{\alpha\sigma}(q)\}$ are termed as
nonergodicity parameters for collective modes. In solving the
Eqns. (\ref{tde:rr})-(\ref{tde:m}), the bare transport
coefficients $L_0(q)$ and $\gamma_0(q)$ are estimated in terms of
those of a hard sphere system, obtained from short time models
\cite{dufty92} of generalized hydrodynamics. Here the
Percus-Yevick \cite{PY} structure factors for the mixture
\cite{lebowitz} is used. The present model predicts an
ergodicity-nonergodicity transition at a critical packing
\cite{jsp2}. In the non-ergodic state the long time limits of the
correlations of $\{\rho,c\}$ fields are nonzero.
%


Next we turn to the tagged particle correlations $\psi_s(q,t)$ for
the two species $s=1,2$. The corresponding time evolution
equations ( described in the main text) involve convolutions with
the respective memory functions ${\Delta_\mathrm{mc}^s}(q,t)$ for
$s=1,2$. These memory functions for the tagged particle
correlations has mode coupling contributions similar to that in
the memory functions for the one component fluid obtained in the
adiabatic approximation \cite{bin2}. For both one component or
multi component systems, the mode coupling effect slaves the
single particle dynamics to the collective dynamics in the dense
system. Since the mode-coupling dynamics of collective variables
are {\em dependent} on the mass ratio \cite{uh-jsp,bin1,bin2} of
the mixture, similar dependence occurs for tagged particle
correlations $\psi_s(q,t)$ ($s=1,2$) and hence for the self
diffusion coefficients $D_s$ ($s=1,2$). The collective correlation
contributions ${\cal C}_s$, $s=1,2$ present in the respective
memory functions, are obtained by solving Eqns.
(\ref{tde:rr})-(\ref{tde:m}) outlined above. To compute ${\cal
C}_s$ in terms of the collective correlation functions, we work in
terms of the conserved densities $\alpha,\sigma\in\{\rho,c\}$
rather than the partial densities $\rho_s$ of species $s=1,2$. The
collective quantities ${\cal C}_s$ ($s=1,2$) are obtained in terms
of the $\phi_{\alpha\sigma}$ as
\be \label{calc-hyd} {\cal C}_s(q,t) = \sum_{\alpha\sigma}
\tilde{c}^{(s)}_{\rho \alpha}(q)\tilde{c}^{(s)}_{\rho \sigma}(q)
\sqrt{S_{\alpha \alpha}(q)S_{\sigma \sigma}(q)}~
\phi_{\alpha\sigma}(q,t)~~. \ee
The functions $\tilde{c}^{(s)}_{\alpha\sigma}$ ( for
${\alpha,\sigma}\in\{\rho,c\}$) are defined as
$\tilde{c}^{(s)}_{\rho\alpha}(q)=c_{\rho
\alpha}(q)+{(-1)}^{s}(x_s-1)~c_{\alpha c}(q)$, for $s=1,2$.
The normalized time correlations $\phi_{ss'}$, and direct
correlations $c_{ss'}$ (for $s,s'=1,2$) are easily transformed to
their counterparts $\phi_{\alpha\sigma}$ and $c_{\alpha\sigma}$,
where $\alpha,\sigma\in\{\rho,c\}$. Beyond a ENE transition point
$\psi_s(q,t\rightarrow\infty)=f_s(q)$ (for $s=1,2$) and $f_s$
termed as nonergodicity parameters. The self diffusion constant
goes to zero  beyond the ENE transition.

\vspace*{5cm} }

\end{document}